\documentclass[11pt, a4paper]{article}

\usepackage{pdfpages}
\usepackage{amsmath}
\usepackage{amsthm}
\usepackage{natbib}
\usepackage{graphicx}
\usepackage[left= 2.5cm, right = 2.5cm, top = 3cm, bottom = 3cm]{geometry}

\usepackage{color}

\newtheoremstyle{example}
{3pt} 
{3pt} 
{} 
{0\parindent} 
{\bf}
{:} 
{.5em} 
{} 
\newtheoremstyle{theorem}
{3pt} 
{3pt} 
{\em} 
{0\parindent} 
{\bf}
{:} 
{.5em} 
{} 
\theoremstyle{example}

\theoremstyle{theorem}

\title{Improving the accuracy of likelihood-based inference in meta-analysis and meta-regression}

\author{I.~Kosmidis \\ Department of Statistical Science \\
  University College London \\ London, WC1E 6BT, U.K \\
  \texttt{i.kosmidis@ucl.ac.uk}
  \bigskip \\
  A.~Guolo \\ Department of Statistical Sciences \\ University of
  Padova \\ Via Cesare Battisti, 241, 35121 Padova, Italy \\
  \texttt{annamaria.guolo@unipd.it}
  \bigskip \\
  C.~Varin \\ Department of Environmental Sciences, Informatics and
  Statistics, \\ Ca' Foscari University of Venice \\ Via Torino 150, 30170
  Venezia Mestre, Italy \\ \texttt{cristiano.varin@unive.it}}

\begin{document}

\maketitle

\begin{abstract}
  Random-effects models are frequently used to synthesise information
  from different studies in meta-analysis. While likelihood-based
  inference is attractive both in terms of limiting properties and of
  implementation, its application in random-effects meta-analysis may
  result in misleading conclusions, especially when the number of
  studies is small to moderate. The current paper shows how
  methodology that reduces the asymptotic bias of the maximum
  likelihood estimator of the variance component can also
  substantially improve inference about the mean effect size. The
  results are derived for the more general framework of random-effects
  meta-regression, which allows the mean effect size to vary with
  study-specific covariates.

\end{abstract}

\section{Introduction}
Meta-analysis is a widely applicable approach to combining information
from different studies about a common effect of interest. A popular
framework for accounting for the heterogeneity between studies is the
random-effects specification in \citet{dersimonian:86}. There is ample
evidence that frequentist inference for this specification can result
in misleading conclusions, especially if inference is carried out by
relying on first-order asymptotic arguments in the common setting of
small or moderate number of studies \citep[e.g.,][]{vanhouwelingen:02,
  guolo:15}. The same considerations apply to the random-effects
meta-regression model, which is a direct extension of random-effects
meta-analysis allowing for study-specific covariates. Proposals
presented to account for the finite number of studies include
modification of the limiting distribution of test statistics
\citep{knapp:03}, restricted maximum likelihood \citep{viechtbauer:05}
and second-order asymptotics \citep{guolo:12}. Recently,
\citet{zeng:15} suggested a double resampling approach that
outperforms several alternatives in terms of empirical coverage
probability of confidence intervals for the mean effect size.

The current paper studies the extent of the bias of the maximum
likelihood estimator of the random-effect variance and introduces a
bias-reducing penalized likelihood that yields a substantial
improvement in the estimation of the random-effect variance. The
bias-reducing penalized likelihood is related to the approximate
conditional likelihood of \citet{cox:87} and the restricted maximum
likelihood for inference about the random-effects variance. The order
of the penalty function allows the derivation of a $\chi^2$
approximation of the distribution of the logarithm of the penalized
likelihood ratio statistic, which can be used for inference about the
fixed-effect parameters. Real-data examples and two simulation studies
illustrate the improvement in finite-sample performance against
alternatives from the recent literature.

\section{Random-effects meta-regression and meta-analysis}

Suppose there are $K$ studies about a common effect of interest, each
of them providing pairs of summary measures $(y_i, \hat\sigma_i^2)$,
where $y_i$ is the study-specific estimate of the effect, and
$\hat\sigma_i^2$ is the associated estimation variance
$(i=1, \ldots, K)$. In some situations, the pairs
$(y_i, \hat\sigma_i^2)$ may be accompanied by study-specific
covariates $x_i = (x_{i1}, \ldots, x_{ip})^\top$, which describe the
heterogeneity across studies. In the meta-analysis literature, it is
usually assumed that the within-study variances $\hat\sigma_i^2$ are
estimated well enough to be considered as known and equal to the
values reported in each study. Under this assumption, the
random-effects meta-regression model postulates that
$y_1, \ldots, y_K$ are realizations of random variables
$Y_1, \ldots, Y_K$, respectively, which are independent conditionally
on independent random effects $U_1, \ldots, U_K$, and the conditional
distribution of $Y_i$ given $U_i = u_i$ is
$N(u_i + x_i^\top\beta, \hat\sigma_i^2)$, where $\beta$ is an unknown
$p$-vector of effects.
The random effect $U_i$ is typically assumed to be
distributed according to $N(0, \psi)$, where
$\psi$ accounts for the between-study heterogeneity.

In matrix
notation, and conditionally on $(U_1, \ldots, U_K)^\top = u$, the
random-effects meta-regression model is
\begin{equation}
\label{eq:metaregression}
Y= X \beta + u +  \epsilon,
\end{equation}
where $Y = (Y_1, \ldots, Y_K)^\top$, $X$ is the model matrix of
dimension $K \times p$ with $x_i^\top$ in its $i$th row, and
$\epsilon=(\epsilon_1, \ldots, \epsilon_K)^\top$ is a vector of
independent errors each with a $N(0, \hat\sigma_i^2)$
distribution. Under this specification, the marginal distribution of
$Y$ is multivariate normal with mean $X\beta$ and variance
$\hat\Sigma+ \psi I_{K}$, where $I_K$ is the $K \times K$ identity
matrix and
$\hat\Sigma=\text{diag}(\hat\sigma_1^2, \ldots, \hat\sigma^2_K)$. The
random-effects meta-analysis model is a meta-regression model where
$X$ is a column of ones.

The random-effects meta-regression model is used here as a working
model for theoretical development. In light of the recent criticisms
of the assumption of known within-study variances \citep[see, for
example][]{hoaglin:15}, \S~\ref{sect:simu} and the Supplementary
Material illustrate the good performance of the derived procedures
under more realistic scenarios, where the estimation variances are
directly related to the estimates of the summary measure.

The parameter $\beta$ is naturally estimated by weighted least squares as
\begin{equation}
\label{eq:betahat}
\hat\beta(\psi) = \{ X^\top  W(\psi) X \}^{-1} X^\top W(\psi) Y \, ,
\end{equation}
with $W(\psi)=(\hat\Sigma+ \psi I_K)^{-1}$.  Then, inference about
$\beta$ can be based on the fact that under
model~(\ref{eq:metaregression}), $\hat\beta(\psi)$ has an asymptotic
normal distribution with mean $\beta$ and variance $\left\{X^\top W(\psi)
X\right\}^{-1}$. In this case, the reliability of the associated inferential
procedures critically depends on the availability of an accurate
estimate of the between-study variance $\psi$. A popular choice is the
\citet{dersimonian:86} estimator
$\hat\psi_{\text{DL}}= \max\left\{0, (Q-n+p)/ A \right\}$, where
$Q=(Y-X \hat\beta_{\text{F}})^\top \hat\Sigma^{-1} (Y-X \hat
\beta_{\text{F}})$ is the Cochran statistic, with
$\hat\beta_{\text{F}} = \hat\beta(0)$
and
$A=\text{tr}(\hat\Sigma^{-1}) - \text{tr}\{ (X^\top \hat
\Sigma^{-1}X)^{-1} X^\top\hat\Sigma^{-2}X \}$. \citet{viechtbauer:05}
presents evidence of the loss of efficiency of $\hat\psi_{\text{DL}}$,
which can impact inference; see also \citet{guolo:12}.

Inference about $\beta$ can alternatively be based on the likelihood
function. The log-likelihood function for $\theta=(\beta^\top, \psi)^\top$
in model~(\ref{eq:metaregression}) is
\begin{equation}
\label{eq:loglikelihood}
\ell(\theta) = \frac{1}{2} \log | W(\psi) | - \frac{1}{2} R(\beta)^\top W(\psi) R(\beta),
\end{equation}
where $| W(\psi) |$ denotes the determinant of $W(\psi)$ and
$R(\beta)=y - X \beta$. A calculation of the gradient $s(\theta)$ of
$\ell(\theta)$ shows that the maximum likelihood estimator
$\hat\theta_{\text{ML}}=(\hat\beta_{\text{ML}}^\top,
\hat\psi_{\text{ML}})^\top$ for $\theta$ results from solving the
equations
\begin{equation}
\label{eq:scores}
\begin{cases}
  s_{\beta}(\theta) = X^\top W(\psi) R(\beta) = 0_p \, , \\
  s_{\psi}(\theta) = R^\top(\beta) W(\psi)^2 R(\beta)/2 - \text{tr}\left[W(\psi)
  \} \right]/2 = 0 \, , \\
\end{cases}
\end{equation}
where $0_p$ denotes a $p$-dimensional vector of zeros, and
$s_{\beta}(\theta) = \nabla_\beta \ell(\theta)$ and
$s_{\psi}(\theta) = \partial \ell(\theta)/\partial \psi$, so that
$\hat\beta_{ML} = \hat\beta(\hat\psi_{ML})$. As observed in
\citet{guolo:12} and \citet{zeng:15}, inferential procedures that rely
on first-order approximations of the log-likelihood, e.g.,
likelihood-ratio and Wald statistics, perform poorly when the number
of studies $K$ is small to moderate.

\section{Bias reduction}
\subsection{Bias-reducing penalized likelihood}\label{sect:bias}
\label{sec:bias}
From the results in \citet{kosmidis:09, kosmidis:10}, the first term
in the expansion of the bias function of the maximum likelihood
estimator is found to be
$b(\theta)=\{ 0_p^\top, b_{\psi}(\psi) \}^\top$, where
\begin{equation}
\label{eq:bias_tau2}
b_{\psi}(\psi)=  -\frac{\text{tr}\{ W(\psi) H(\psi)
  \}}{\text{tr}\{W(\psi)^{2}\}}\, ,
\end{equation}
with $H(\psi)=X\{ X^\top W(\psi) X\}^{-1} X^\top W(\psi)$. A sketch
derivation for (\ref{eq:bias_tau2}) is given in the Appendix. In what
follows $b(\theta)$ is called the first-order bias.

The non-zero entries of $W(\psi)$ and the diagonal entries of
$H(\psi)$ are all necessarily positive, so the maximum likelihood
estimator of $\psi$ is subject to downward bias, which, as also noted
in \citet{viechtbauer:05}, affects inference about $\beta$, by
over-estimating the non-zero entries of $W(\psi)$, and hence
over-estimating the information matrix
\begin{equation}
\label{eq:fisherinfo}
F(\theta) =  -E_\theta\left\{\frac{\partial^2 \ell(\theta)}{\partial \theta \partial
\theta^\top} \right\} =
 \left[
\begin{array}{cc}
X^\top W(\psi) X & 0_p \\
0_p^\top & \frac{1}{2}\text{tr}\left\{W(\psi)^2\right\}
\end{array}
\right] \,.
\end{equation}
This over-estimation of $F(\theta)$ can result in hypothesis tests
with large Type I error and confidence intervals or regions with
actual coverage appreciably lower than the nominal level.

An estimator that corrects for the first-order bias of
$\hat\theta_{\text{ML}}$ results from solving the adjusted score
equations $s^*(\theta) = s(\theta) - F(\theta)b(\theta) = 0_{p+1}$
\citep{firth:93, kosmidis:09}. Substituting (\ref{eq:scores}),
(\ref{eq:bias_tau2}) and (\ref{eq:fisherinfo}) in the expression for
$s^*(\theta)$ gives that the adjusted score functions for $\beta$ and
$\psi$ are $s^*_{\beta}(\theta) = s_{\beta}(\theta)$ and
\begin{equation}
\label{eq:adjustedpsi}
  s^*_{\psi}(\theta) = R^\top(\beta) W(\psi)^2 R(\beta)/2 -
  \text{tr}\left[W(\psi) \{ I_K - H(\psi) \} \right]/2 = 0 \, ,
\end{equation}
respectively. The expression for the differential of the
log-determinant can be used to show that $s^*_{\beta}(\theta)$ and
$s^*_{\psi}(\theta)$ are the derivatives of the penalized
log-likelihood function
\begin{equation}
\label{eq:penloglik}
\ell^*(\theta) = \ell(\theta) - \frac{1}{2} \log \left| F_{(\beta\beta)}(\psi) \right|\, ,
\end{equation}
where $\ell(\theta)$ is as in (\ref{eq:loglikelihood}),
$F_{(\beta\beta)}(\psi) = X^\top W(\psi) X$ is the $\beta$-block of the
information matrix $F(\theta)$, and $\left|\cdot\right|$ denotes
determinant, so the solution of the adjusted score equations is the
maximum penalized likelihood estimator $\hat\theta_{\text{MPL}}$.

For $\beta=\hat\beta(\psi)$, expression~(\ref{eq:penloglik}) reduces
to both the logarithm of the approximate conditional likelihood of
\citet{cox:87} for inference about $\psi$, when $\beta$ is treated as
a nuisance component, and to the restricted log-likelihood function of
\citet{harville:77}. Hence, maximising the bias-reducing penalized
log-likelihood (\ref{eq:penloglik}) is equivalent to calculating the
maximum restricted likelihood estimator for $\psi$. The latter
estimator was originally constructed to reduce underestimation of
variance components in finite samples as a consequence of failing to
account for the degrees of freedom that are involved in the estimation
of the fixed effects $\beta$. \citet{smyth:96} and \citet{stern:00}
have shown the equivalence of the restricted log-likelihood with
approximate conditional likelihood in the more general context of
inference about variance components in normal linear mixed models.

\subsection{Estimation}
Given a starting value $\psi^{(0)}$ for $\psi$, the following
iterative process has a stationary point that maximizes
(\ref{eq:penloglik}). At the $j$th iteration $(j = 1, 2, \ldots )$, a
new candidate value $\beta^{(j+1)}$ for $\beta$ is obtained as the
weighted least squares estimator (\ref{eq:betahat}) at
$\psi=\psi^{(j)}$; a candidate value for $\psi^{(j + 1)}$ is then
computed by a line search for solving the adjusted score equation
(\ref{eq:adjustedpsi}) evaluated at $\beta=\beta^{(j+1)}$. The
iteration is repeated until either the candidate values do not change
across iterations or the adjusted score functions are sufficiently
close to zero.

\subsection{Penalized likelihood inference}

The profile penalized likelihood function can be used to construct
confidence intervals and regions, and carry out hypothesis tests for
$\beta$. If $\beta = (\gamma^\top, \lambda^\top)^\top$, and
$\hat\lambda_{\text{MPL}, \gamma}$ and $\hat\psi_{\text{MPL}, \gamma}$
are the estimators of $\lambda$ and $\psi$, respectively, from
maximising (\ref{eq:penloglik}) for fixed $\gamma$, then the logarithm
of the penalized likelihood ratio statistic
$2\{\ell^*(\hat\gamma_{\text{MPL}}, \hat\lambda_{\text{MPL}},
\hat\psi_{\text{MPL}}) - \ell^*(\gamma, \hat\lambda_{\text{MPL},
  \gamma}, \hat\psi_{\text{MPL}, \gamma})\}$ has the usual limiting
$\chi^2_q$ distribution, where $q = \dim(\gamma)$. To derive this
limiting result, note that the adjustment to the scores in
(\ref{eq:scores}) is additive and $O(1)$, so the extra terms depending
on it and its derivatives in the asymptotic expansion of the penalized
likelihood disappear as information increases.

The impact of using the penalized likelihood for estimation and
inference in random-effects meta-analysis and meta-regression is more
profound for a small to moderate number of studies. As the number of
studies increases, the log-likelihood derivatives dominate the
bias-reducing adjustment in (\ref{eq:adjustedpsi}) in terms of
asymptotic order. As a result, inference based on the penalized likelihood
becomes indistinguishable from likelihood inference.

In \S~\ref{sect:simu}, the performance of penalized likelihood
inference is compared with that of alternative methods under the
random-effects meta-analysis model (\ref{eq:metaregression}), and
under a more realistic model for individual-within-study data.

\section{Simulation studies}\label{sect:simu}

\subsection{Random-effects meta-analysis}
\label{sec:simu1}
The simulation studies under the random-effects meta-analysis model
(\ref{eq:metaregression}) are performed using the design in
\citet{brockwell:01}.  Specifically, the study-specific effects $y_i$
are simulated from the random-effect meta-analysis with true effect
$\beta=0.5$ and variance $\hat\sigma^2_i + \psi$, where
$\hat\sigma^2_i$ are independently generated from a $\chi^2_1$
distribution multiplied by $0.25$ and then restricted to the interval
$(0.009, 0.6)$. The between-study variance $\psi$ ranges from $0$ to
$0.1$ and the number of studies $K$ from $5$ to $200$. For each
combination of $\psi$ and $K$ considered, $10\, 000$ data sets are
simulated using the same initial state for the random number
generator.

\begin{figure}
\begin{center}
  \includegraphics[width = 0.8\textwidth]{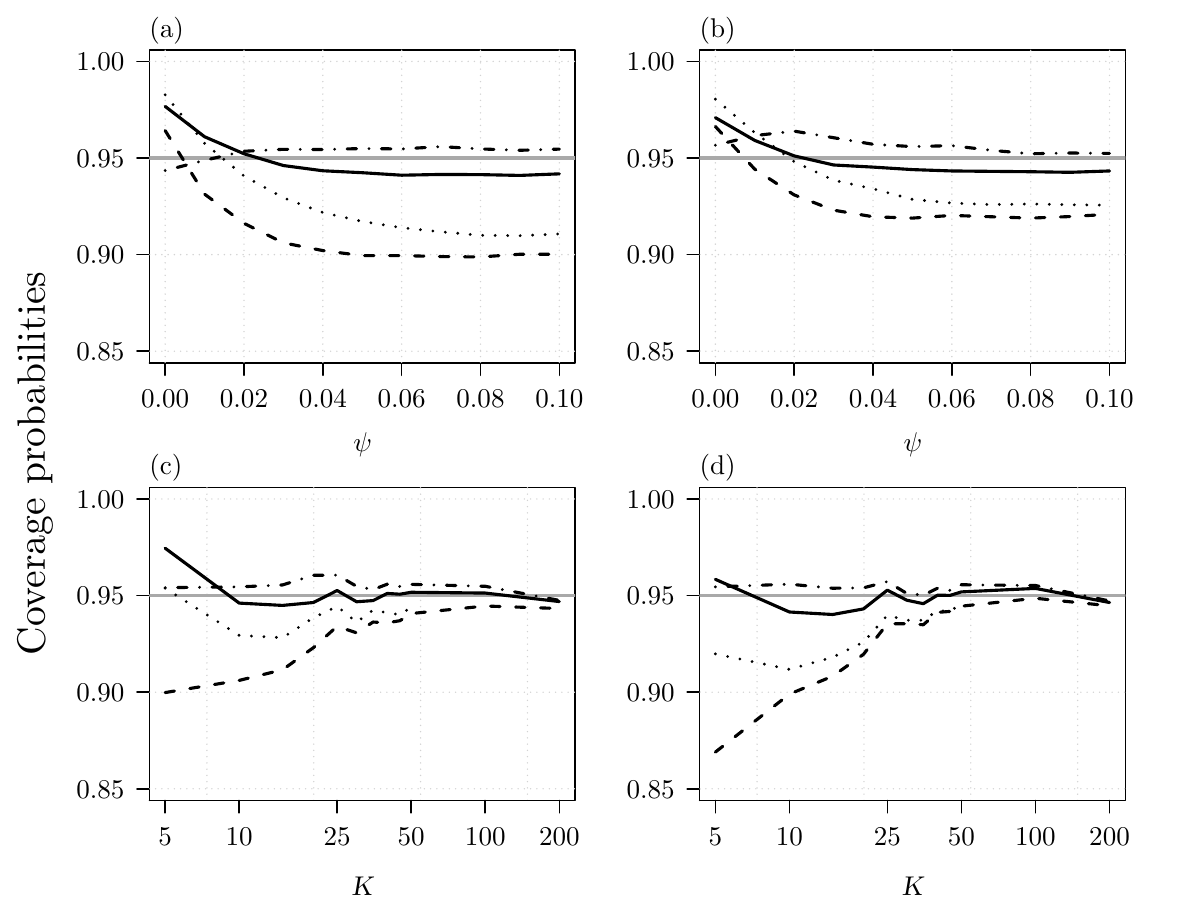}
\end{center}
\caption{Empirical coverage probabilities of two-sided confidence
  intervals for $\beta$ for increasing $\psi$, when (a) $K=10$ and (b)
  $K=20$, and for increasing $K$ (in log scale) when (c) $\psi=0.03$
  and (d) $\psi=0.07$. The curves correspond to profile penalized
  likelihood (solid), DerSimonian \& Laird method (dashed), Zeng \&
  Lin double resampling (dotted; available only for $K \le 50$), and
  Skovgaard's statistic (dotted-dashed). The grey horizontal line is
  the target 95\% nominal level.}\label{fig:coverages1}
\end{figure}

\citet[Section 5]{zeng:15} show that their double resampling approach
outperforms several existing methods in terms of the empirical
coverage probabilities of confidence intervals for $\beta$ at nominal
level 95\%. The methods considered in \citet{zeng:15} include profile
likelihood \citep{hardy:96}, modified DerSimonian \& Laird
\citep[see][]{sidik:02,knapp:03,copas:03},
quantile approximation \citep{jackson:09} and the approach described
in \citet{henmi:10}. The present simulation study takes advantage of
these previous simulation results, and Figure~\ref{fig:coverages1}
compares the performance of double resampling with that of the profile
penalized likelihood. In order to avoid long computing
times, empirical coverage for double-resampling has been calculated
only for $K \le 50$.

The profile penalized likelihood confidence interval has empirical
coverage that is appreciably closer to the nominal level than double
resampling.

Figure \ref{fig:coverages1} also includes results for two alternative
confidence intervals. The first uses the classical DerSimonian \&
Laird estimator $\hat\beta(\psi_{\text{DL}})$ and its estimated
variance $1/\sum_{i = 1}^K 1/(\hat\sigma_i +
\hat\psi_{\text{DL}})$. Not surprisingly, the empirical coverage of
this confidence interval is grossly smaller than the nominal
confidence level. The second interval is used for reference and
results from the numerical inversion of Skovgaard's statistic, which
is designed to produce second-order accurate p-values for tests on the
mean effect size \citep{guolo:12, guolo:12b}. The profile penalized
likelihood interval has comparable performance to that based on
Skovgaard's statistic, with the latter having empirical coverage
slightly closer to the nominal level for a wider range of values for
$\psi$.
In general, though, the numerical inversion of Skovgaard's statistic
can be unstable due to the discontinuity of the statistic around the
maximum likelihood estimator. In contrast, the calculation of profile
penalized likelihood intervals is not prone to such
instabilities. The penalized likelihood also results in
a bias-reduced estimator of $\psi$, whose reliable estimation is often
of interest in medical studies \citep{veroniki:16}.

\subsection{Standardized mean differences from two-arm studies}
\label{sec:simu2}

\begin{figure}
  \begin{center}
    \includegraphics[width = 0.8\textwidth]{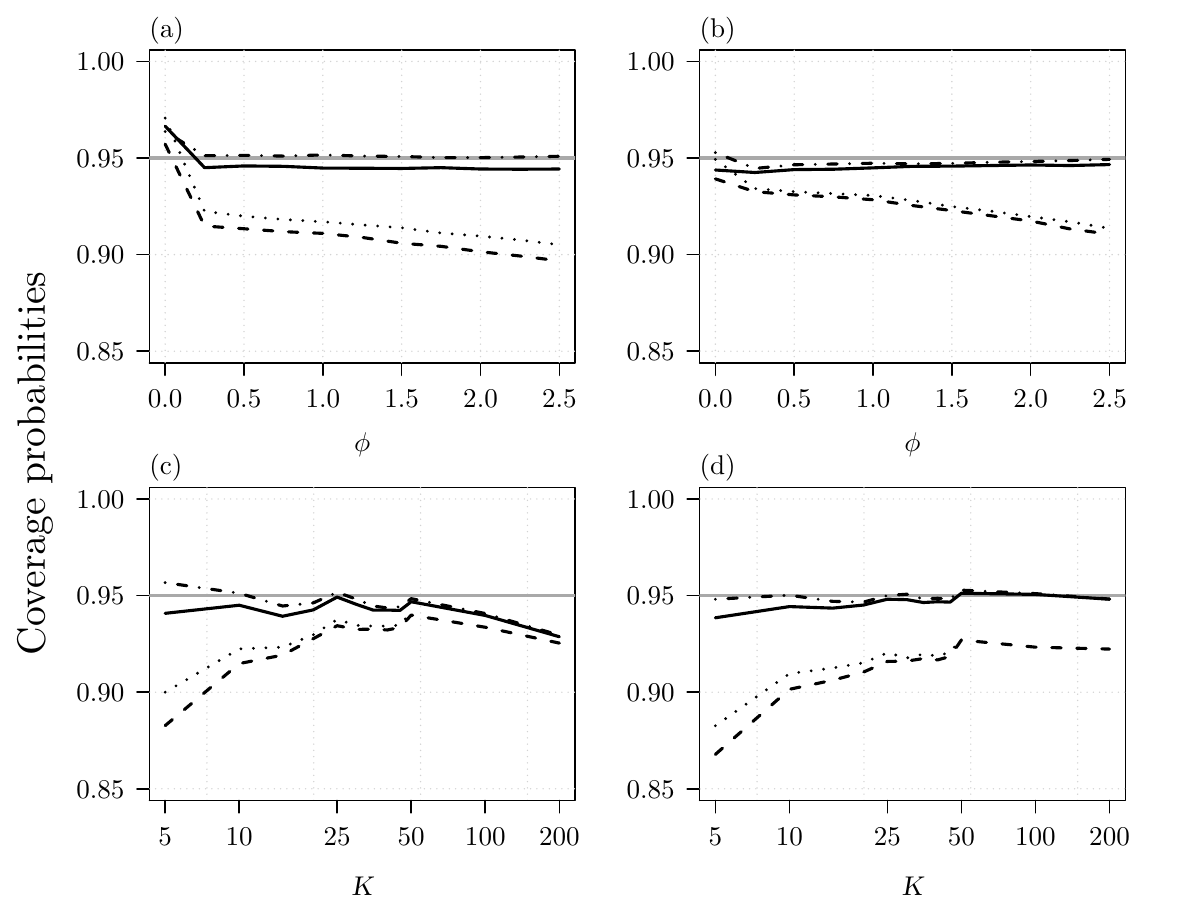}
  \end{center}
  \caption{Empirical coverage probabilities of two-sided confidence
  intervals for $\delta$ for increasing $\phi$, when (a) $K=10$ and
  (b) $K=35$, and for increasing $K$ (in log scale) when (c)
  $\phi=0.25$ and (d) $\phi=2$. The curves correspond to profile
  penalized likelihood (solid), DerSimonian \& Laird method (dashed),
  Zeng \& Lin double resampling (dotted), and Skovgaard's statistic
  (dotted-dashed). The grey horizontal line is the target 95\% nominal
  level.}\label{fig:coverages2}
\end{figure}

The profile penalized likelihood and all other methods in
Figure~\ref{fig:coverages1} have been developed under the validity of
the random-effects meta-analysis model. This assumption may be
unrealistic, especially in settings where the estimation variances are
directly related to the summary measure \citep{hoaglin:15}. Here, we
examine the performance of the methods under an alternative
specification of the data generating process, where the study-specific
effects and their variances are calculated by simulating
individual-within-study data.  Specifically, we assume that the $i$th
study consists of two arms with $n_i$ individuals each, and that
$n_1, \ldots, n_K$ are independent uniform draws from the integers
$\{30, 31, \ldots, 100\}$. Then, conditionally on a random effect
$\alpha_i \sim N(0, \phi)$, we assume that the observation $z_{i, rj}$
for the $j$th individual in the $r$th arm is the realisation of a
$N(\mu + I_r(\delta + \alpha_i)\sigma, \sigma^2)$ random variable,
where $I_1 = 0$ and $I_2 = 1$. The difference between the marginal
variances of the arms increases with $\phi$. The true effects are set
to $\mu = 0$, $\sigma = 1$ and $\delta = -2$. The study-specific
effect of interest is $\delta,$ estimated using the standardized mean
difference $y_i = J_i(\bar{z}_{i,2} - \bar{z}_{i,1})/s_i$, where
$s_i^2$ is the pooled variance from the two arms of the $i$th study,
and $J_i = 1 - 3/\{8(n_i - 1)- 1\}$ is the Hedges correction
\citep[see, {e.g.},][Chapter 4]{borenstein:09}. The corresponding
estimated variance for $y_i$ is
$\hat{\sigma}_i^2 = 2J_i/n_i + J_iy_i^2/(4n_i)$, which is a quadratic
function of $y_i$. The between-study variance $\phi$ ranges from $0$
to $2.5$ and the number of studies $K$ from $5$ to $200$. For each
combination of $\phi$ and $K$ considered, $10\, 000$ data sets are
simulated using the same initial state for the random number
generator.

Figure~\ref{fig:coverages2} shows the empirical coverage of the
confidence intervals for $\delta$ based on the methods that were
examined in Figure~\ref{fig:coverages1}. Empirical coverage for
double-resampling has again been calculated only for $K \le 50$. The
good performance of the profile penalized likelihood interval and the
interval based on the Skovgaard's statistic persists for small and
moderate number of studies, even under the alternative data generating
process. The performance of the intervals based on the
DerSimonian~\&~Laird estimator and double resampling is, again, poor.

Figure~\ref{fig:coverages2} also illustrates the effect of increasing
the number of studies under the alternative specification of the data
generating process. As the number of studies increases, the inadequacy
of the assumptions of the working random-effects meta-regression model
becomes more notable. Model mis-specification will eventually result
in loss of coverage for all methods examined here, including the
intervals based on profile penalized likelihood and the Skovgaard's
statistic.

The Supplementary Material provides the full results from this study
and two other simulation studies, where the summary measures are
log-odds-ratios from a case-control study.

\section{Case study: meat consumption data}\label{sec:illustration}

\citet{larsson:14} investigate the association between meat
consumption and relative risk of all-cause mortality. The data include
$16$ prospective studies, eight of which are about unprocessed red
meat consumption and eight about processed meat consumption. We
consider meta-regression with a covariate taking value $1$ for
processed red mean and 0 for unprocessed.  The DerSimonian \& Laird
estimate of $\psi$ is $\hat\psi_{\text{DL}} = 0.57 \times 10^{-2}$,
the maximum likelihood estimate is
$\hat\psi_{\text{ML}} = 0.85 \times 10^{-2}$ 
and the maximum penalized likelihood estimate is the largest with
$\hat\psi_{\text{MPL}} = 1.18 \times 10^{-2}$. The estimates of
$\beta$ are $\hat\beta_{\text{ML}} = (0.10,
0.11)^\top$,
$\hat\beta_{\text{MPL}} = (0.09, 0.11)^\top$ and $\hat\beta_{\text{DL}}
= (0.11,
0.10)^\top$, where the first element in each vector corresponds to the
intercept and the second to meat consumption.

The DerSimonian~\&~Laird method indicates some evidence for a higher
risk associated to the consumption of red processed meat with a
p-value of $0.027$. In contrast, the penalized likelihood ratio and
Skovgaard's statistic suggest that there is rather weak evidence for
higher risk, with p-values of $0.066$ and
$0.073$, respectively.

The Supplementary Material contains a simulation study under the
maximum likelihood fit that illustrates that the maximum likelihood
estimator of $\psi$ is negatively biased. The other estimators almost
fully compensate for that bias, but $\hat\psi_{\text{MPL}}$ is
appreciably more efficient than $\hat\psi_{\text{DL}}$. The simulation
study therein is also used to illustrate the good performance of the
penalized likelihood ratio test in terms of size.

\section{Supplementary material}
The Supplementary Material provides R \citep{R:15} code to reproduce
the case study in \S~\ref{sec:illustration}, and another analysis.
The full results of the simulations are provided including the
performance of confidence intervals based on alternative methods.

\section*{Acknowledgements}
This work was partially supported by grants `IRIDE', Ca' Foscari
University of Venice and `Progetti di Ricerca di Ateneo 2015',
University of Padova. The authors thank David Firth for discussions
and feedback on an early version of the paper. The authors are also
grateful to two anonymous Referees, the Associate Editor and the
Editor, for helpful comments and suggestions, and thankful to Sophia
Kyriakou for bringing to their attention some inconsequential typos
that have been fixed in the current version. In particular, expression
(\ref{eq:loglikelihood}) has $\ell(\theta) = \frac{1}{2} \, ...$
instead of $\ell(\theta) = - \frac{1}{2} \, ...$, and the middle
expressions in the equations for $s_{\psi}(\theta)$ and
$s^*_{\psi}(\theta)$ in (\ref{eq:scores}) and (\ref{eq:adjustedpsi}),
respectively, have been multiplied by a factor of $1/2$.

\appendix
\section*{Appendix A: derivation of the expression for the first-order bias}
The first-order bias of $\hat\theta_{(ML)}$ has the form
$b(\theta) = - \left\{F(\theta)\right\}^{-1}A(\theta)$
\citep{kosmidis:10}, where $A(\theta)$ has components
\[
A_t(\theta) = -\frac{1}{2} \text{tr}
\left[\left\{F(\theta)\right\}^{-1}\left\{ P_t(\theta) + Q_t(\theta)
  \right\} \right] \quad (t = 1, \ldots, p + 1)\, .
\]
There,
$P_t(\theta) = E_\theta\{ s(\theta) s(\theta)^\top
s_t(\theta)\}$
and $Q_t(\theta) = E_\theta\{-I(\theta) s_t(\theta)\}$.

The model assumptions imply that
$E_\theta\{R_i(\beta)^m\}$ is $0$ if $m$ is odd and
$(m - 1)!!/w_i(\psi)^{m/2}$ if $m$ is even, where
$w_i(\psi) = 1/(\hat\sigma_i^2 + \psi)$, and $(m - 1)!!$ denotes the
double factorial of $m - 1$ $(m = 1, 2, \ldots; i =1, \ldots,
K)$. Direct matrix calculations give
\[
  P_t(\theta) + Q_t(\theta)  = 0_{(p + 1) \times (p + 1)} \quad (t = 1, \ldots, p) \,; \quad
  P_{p + 1}(\theta) + Q_{p + 1}(\theta) = \left[
  \begin{array}{cc}
    X^\top W(\psi)^2 X & 0_p \\
    0_p^\top & 0
  \end{array}
\right] \, ,
\]
where $0_{p \times p}$ is the $p \times p$ zero matrix. So,
$A_t(\theta) = 0$ for $t \in \{1, \ldots, p\}$ and
\[
A_{p + 1}(\theta) = \text{tr}\left[ \left\{X^\top W(\psi)
    X\right\}^{-1}X^\top W(\psi)^2 X\right] =
\text{tr}\left\{W(\psi)H(\psi)\right\}\,.
\]
Inserting the expressions
for the components of $A(\theta)$ into the expression for $b(\theta)$
gives $b(\theta)=\{ 0_p^\top, b_{\psi}(\psi) \}^\top$, where
$b_{\psi}(\psi)$ is as in (\ref{eq:bias_tau2}).



\section*{Supplementary material}

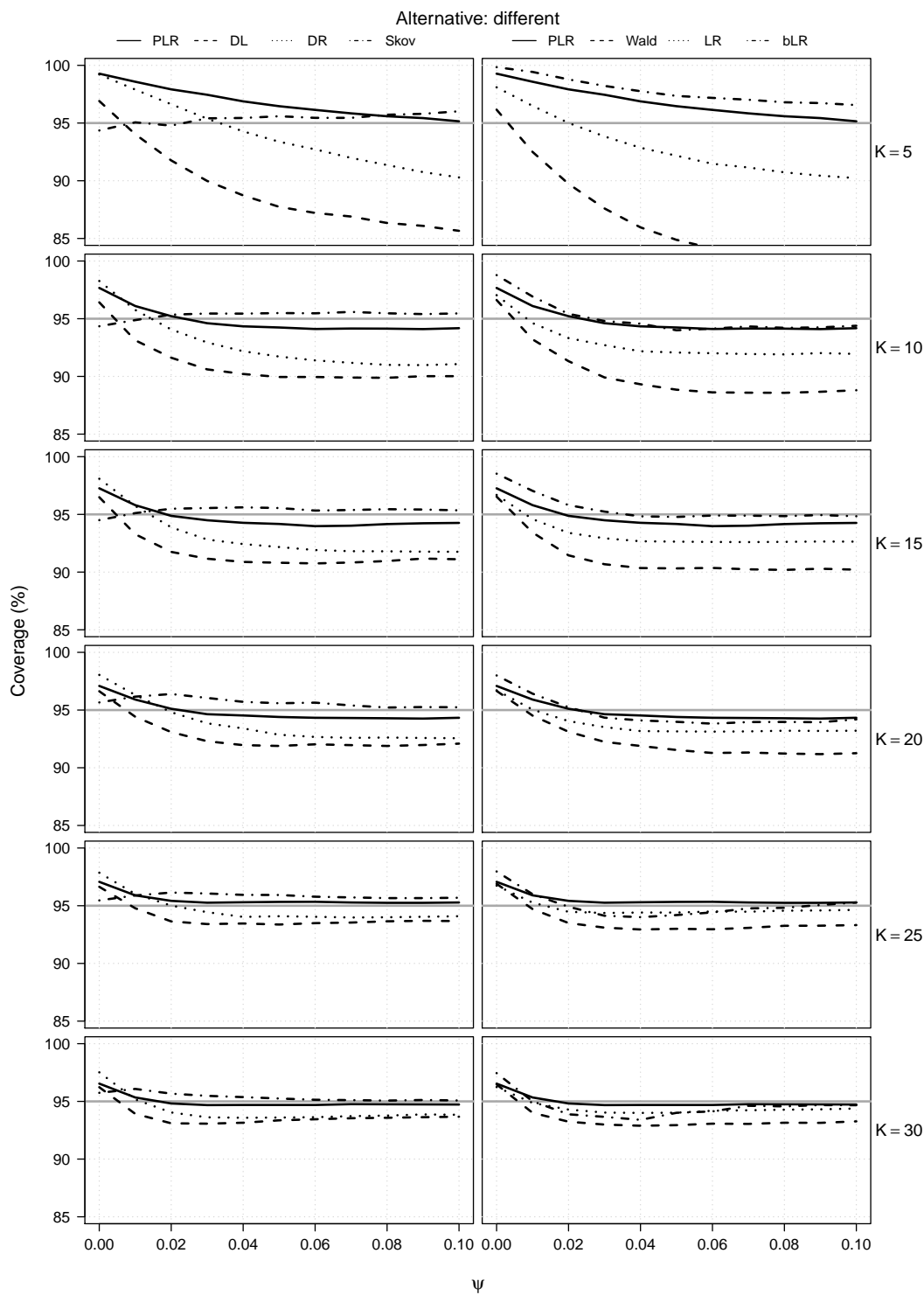

Figure 4: Empirical coverage probabilities of two-sided confidence intervals for random-effects meta-analysis. The empirical coverage is calculated for increasing values of $\psi$ and for $K \in \{5, 10, 15, 20, 25, 30\}$. The curves correspond to the proposed penalized likelihood method (PLR), the DerSimonian & Laird method (DL; left column), the Zeng & Lin double resampling method (DR; left column), the Skovgaard statistic (Skov; left column), the Wald statistic (Wald; right column), the likelihood-ratio statistic (LR; right column), and the Bartlett-corrected likelihood-ratio statistic (bLR; right column). The grey horizontal line is the target 95% nominal level.

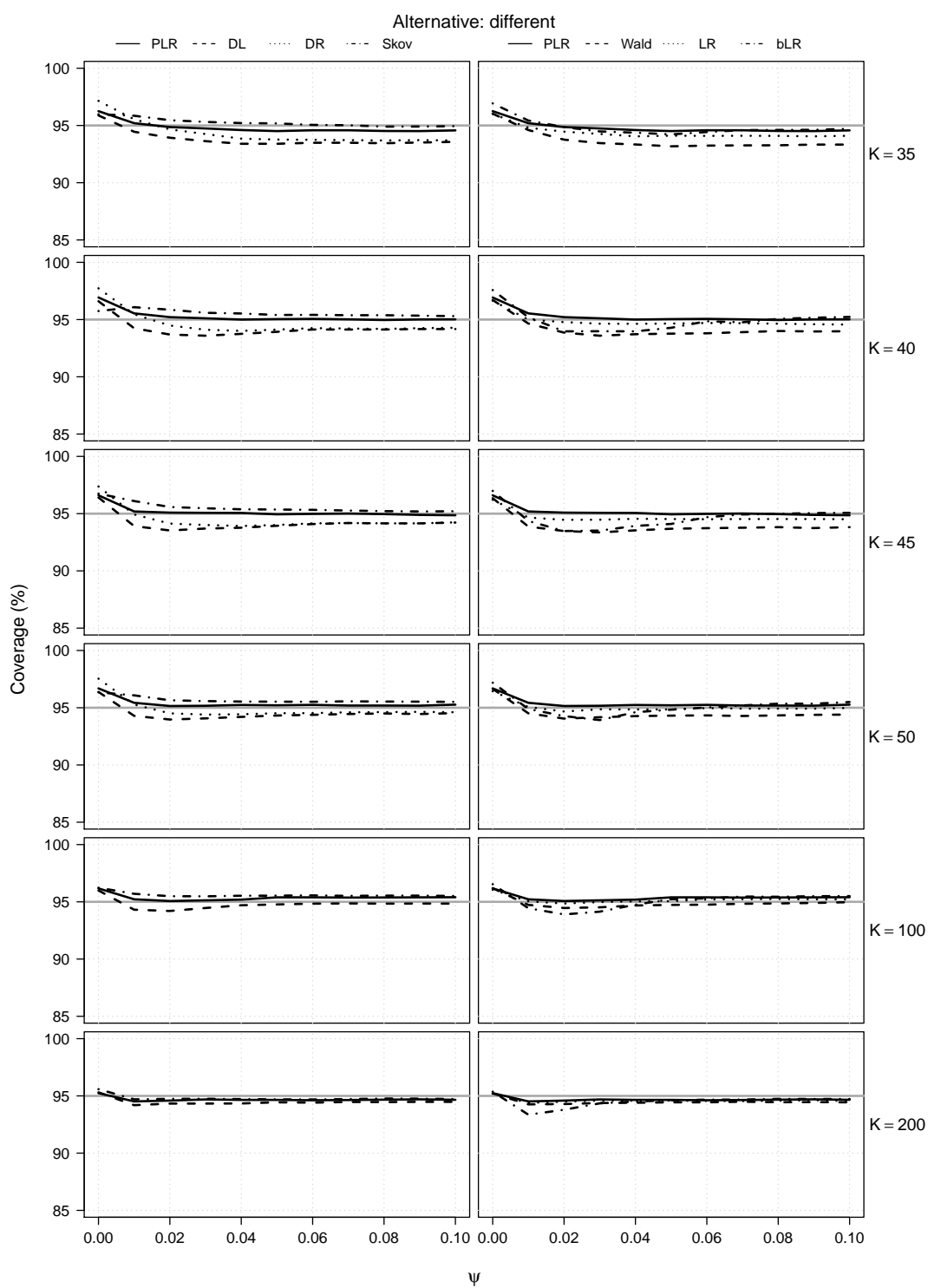

Figure 5: Empirical coverage probabilities of two-sided confidence intervals for random-effects meta-analysis. The empirical coverage is calculated for increasing values of $\psi$ and for $K \in \{35, 40, 45, 50, 100, 200\}$. The curves correspond to the proposed penalized likelihood method (PLR), the DerSimonian & Laird method (DL; left column), the Zeng & Lin double resampling method (DR; left column), the Skovgaard statistic (Skov; left column), the Wald statistic (Wald; right column), the likelihood-ratio statistic (LR; right column), and the Bartlett-corrected likelihood-ratio statistic (bLR; right column). The grey horizontal line is the target 95% nominal level.

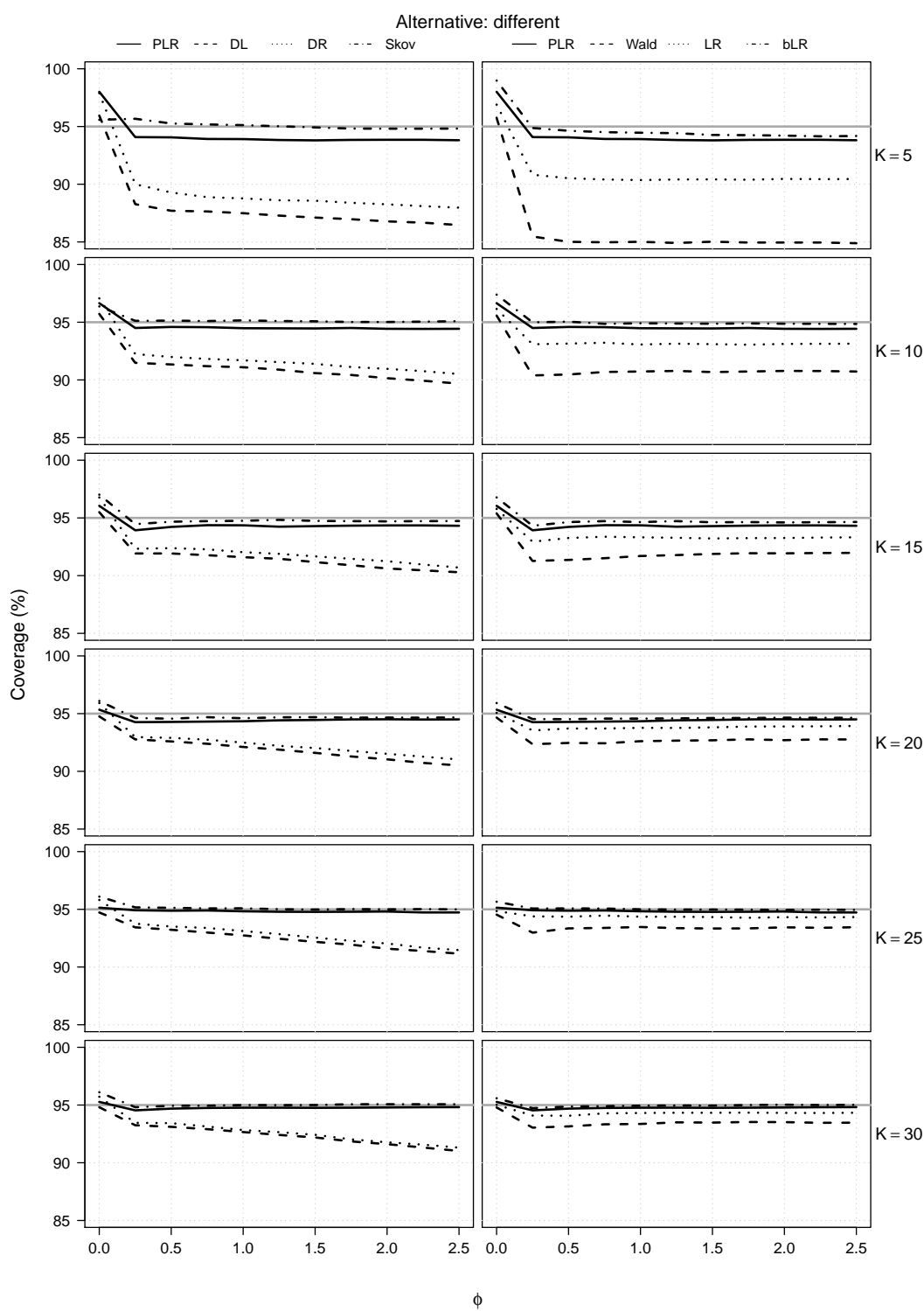

Figure 6: Empirical coverage probabilities of two-sided confidence intervals for standardized mean differences from two-arm studies. The empirical coverage is calculated for increasing values of $\phi$ and for $K \in \{5, 10, 15, 20, 25, 30\}$. The curves correspond to the proposed penalized likelihood method (PLR), the DerSimonian & Laird method (DL; left column), the Zeng & Lin double resampling method (DR; left column), the Skovgaard statistic (Skov; left column), the Wald statistic (Wald; right column), the likelihood-ratio statistic (LR; right column), and the Bartlett-corrected likelihood-ratio statistic (bLR; right column). The grey horizontal line is the target 95% nominal level.

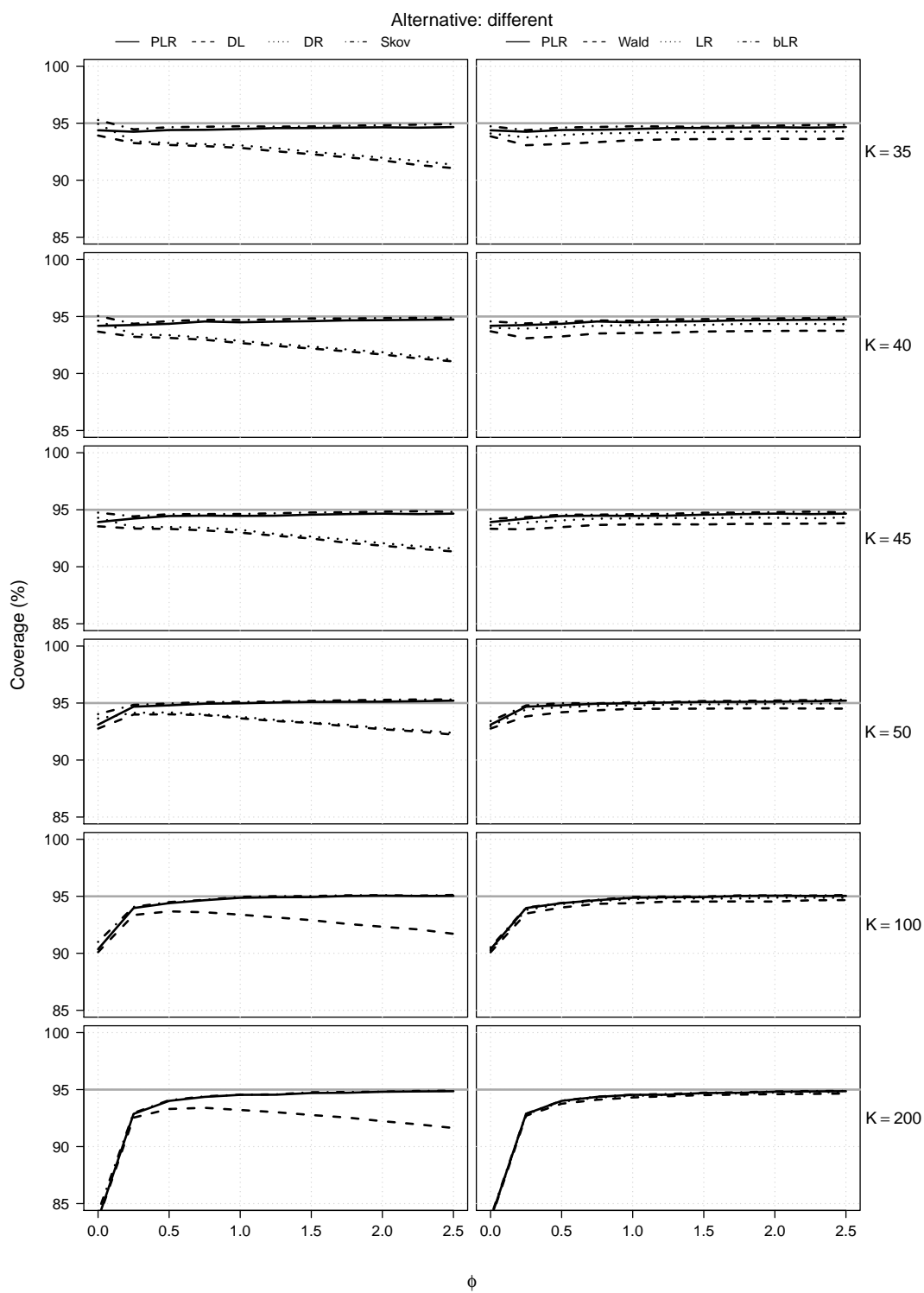

Figure 7: Empirical coverage probabilities of two-sided confidence intervals for standardized mean differences from two-arm studies. The empirical coverage is calculated for increasing values of $\phi$ and for $K \in \{35, 40, 45, 50, 100, 200\}$. The curves correspond to the proposed penalized likelihood method (PLR), the DerSimonian & Laird method (DL; left column), the Zeng & Lin double resampling method (DR; left column), the Skovgaard statistic (Skov; left column), the Wald statistic (Wald; right column), the likelihood-ratio statistic (LR; right column), and the Bartlett-corrected likelihood-ratio statistic (bLR; right column). The grey horizontal line is the target 95% nominal level.

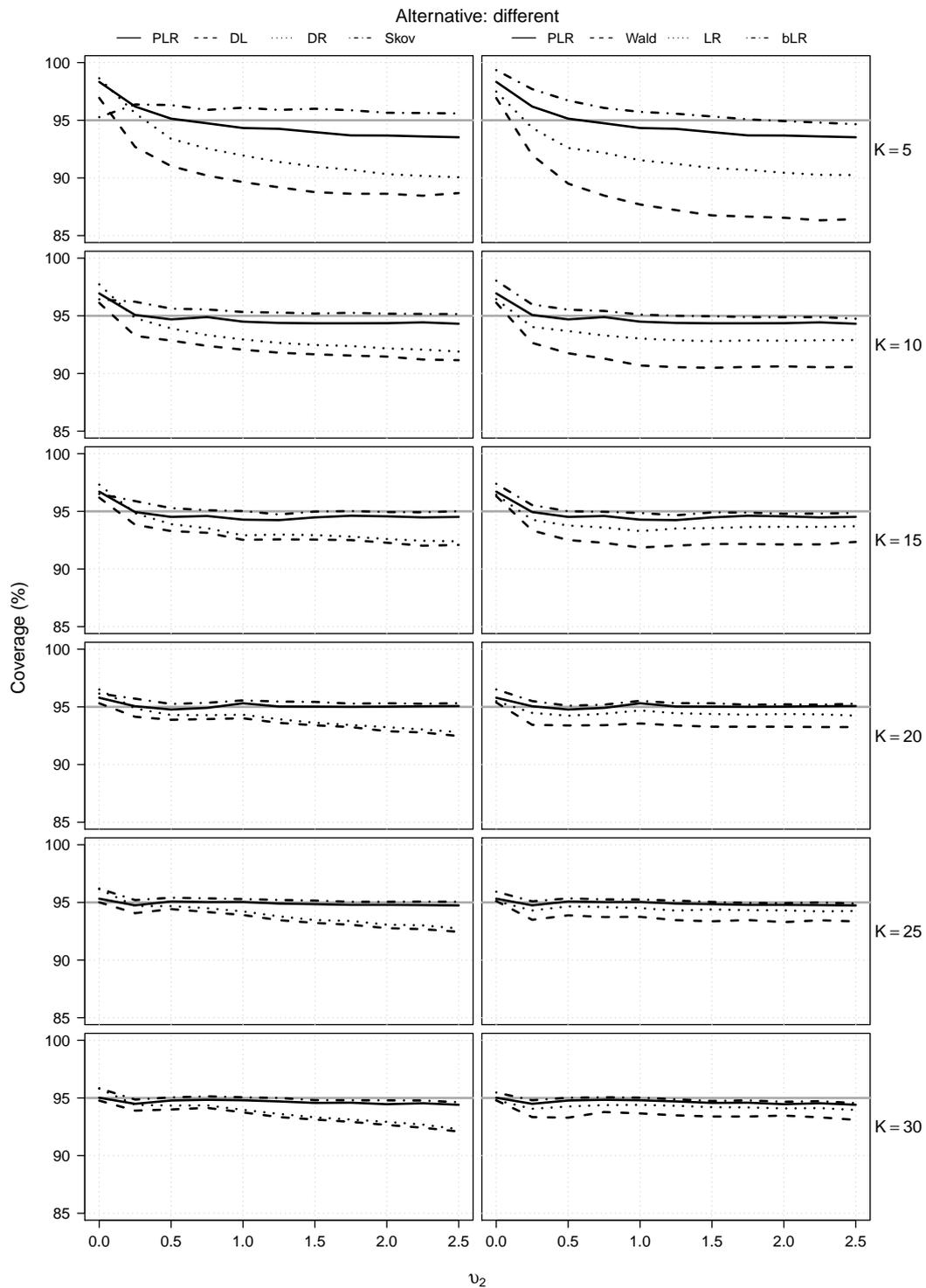

Figure 8: Empirical coverage probabilities of two-sided confidence intervals for log-odds ratios from a case-control study design with moderate effects. The empirical coverage is calculated for increasing values of $v_2$ and for $K \in \{5, 10, 15, 20, 25, 30\}$. The curves correspond to the proposed penalized likelihood method (PLR), the DerSimonian & Laird method (DL; left column), the Zeng & Lin double resampling method (DR; left column), the Skovgaard statistic (Skov; left column), the Wald statistic (Wald; right column), the likelihood-ratio statistic (LR; right column), and the Bartlett-corrected likelihood-ratio statistic (bLR; right column). The grey horizontal line is the target 95% nominal level.

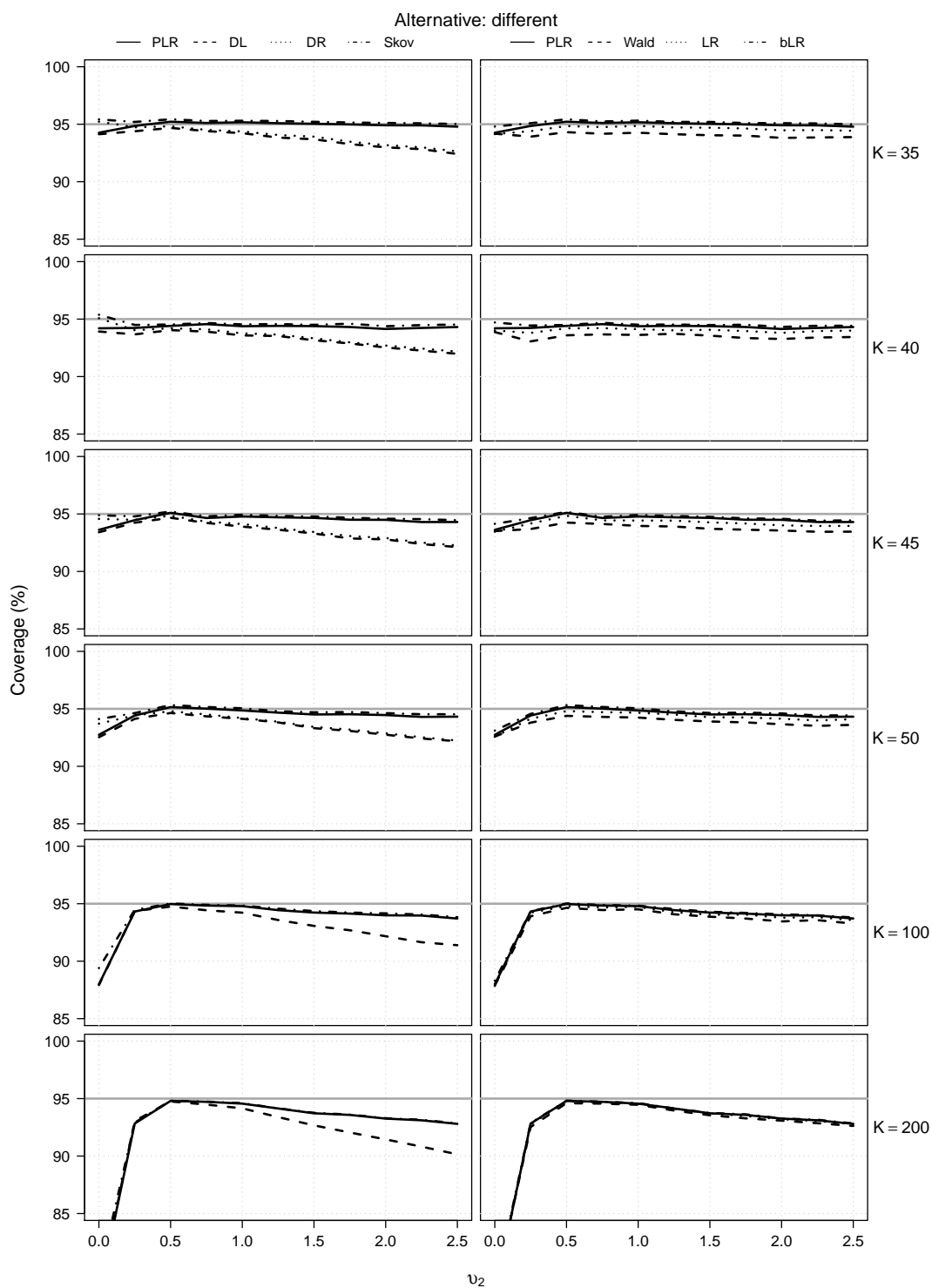

Figure 9: Empirical coverage probabilities of two-sided confidence intervals for log-odds ratios from a case-control study design with moderate effects. The empirical coverage is calculated for increasing values of $v_2$ and for $K \in \{35, 40, 45, 50, 100, 200\}$. The curves correspond to the proposed penalized likelihood method (PLR), the DerSimonian & Laird method (DL; left column), the Zeng & Lin double resampling method (DR; left column), the Skovgaard statistic (Skov; left column), the Wald statistic (Wald; right column), the likelihood-ratio statistic (LR; right column), and the Bartlett-corrected likelihood-ratio statistic (bLR; right column). The grey horizontal line is the target 95% nominal level.

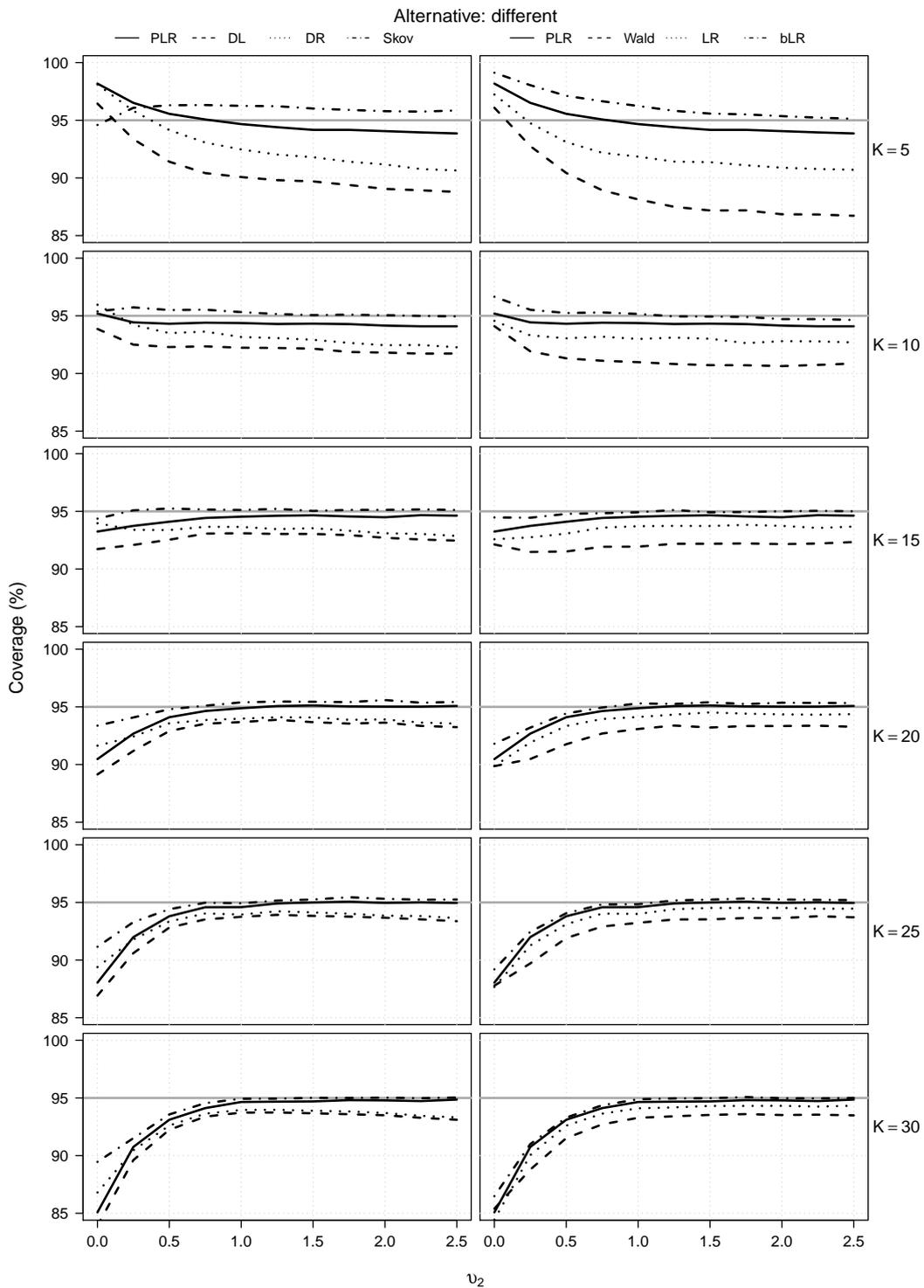

Figure 10: Empirical coverage probabilities of two-sided confidence intervals for log-odds ratios from a case-control study design with larger effects. The empirical coverage is calculated for increasing values of $v_2$ and for $K \in \{5, 10, 15, 20, 25, 30\}$. The curves correspond to the proposed penalized likelihood method (PLR), the DerSimonian & Laird method (DL; left column), the Zeng & Lin double resampling method (DR; left column), the Skovgaard statistic (Skov; left column), the Wald statistic (Wald; right column), the likelihood-ratio statistic (LR; right column), and the Bartlett-corrected likelihood-ratio statistic (bLR; right column). The grey horizontal line is the target 95% nominal level.

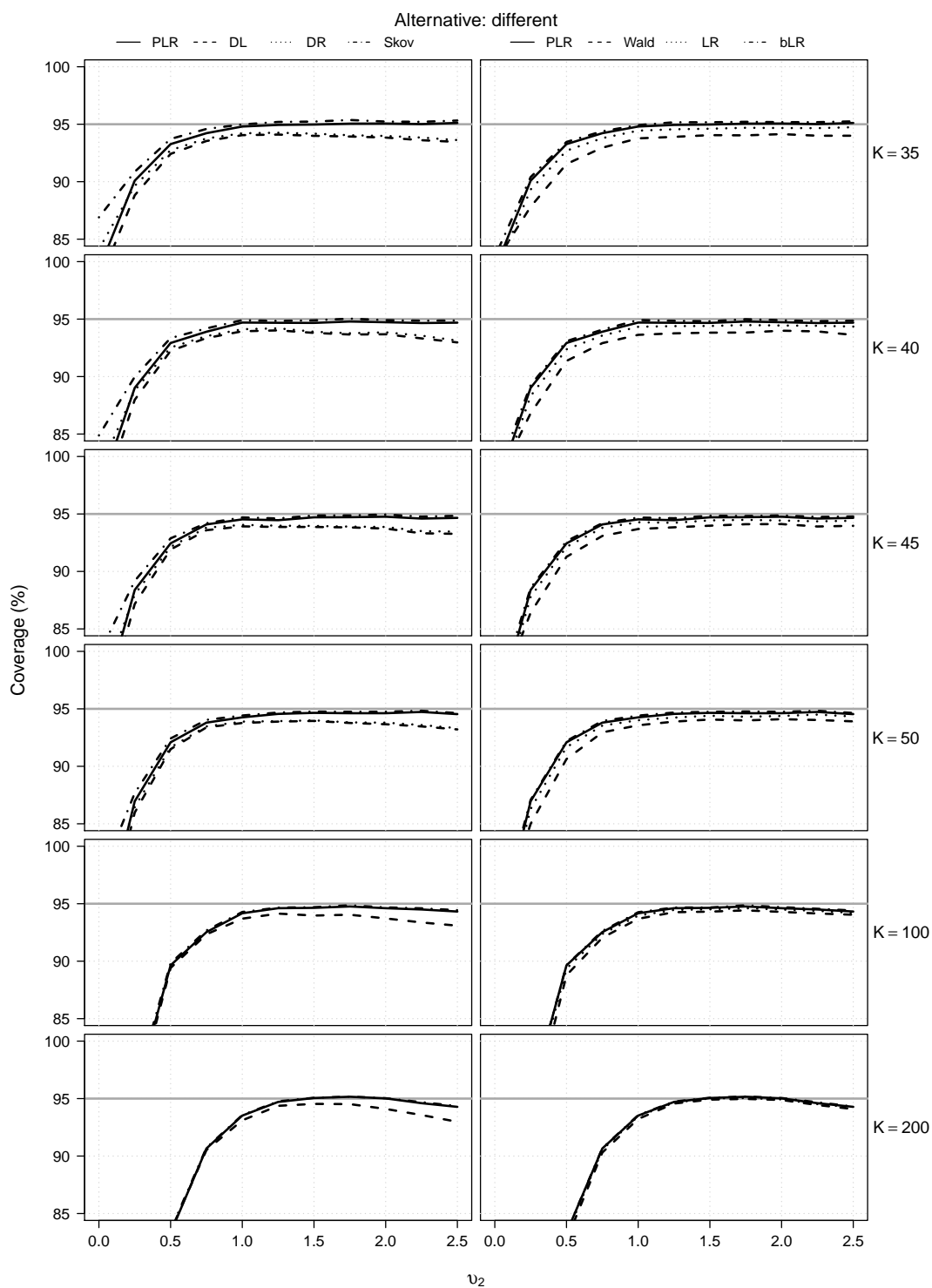

Figure 11: Empirical coverage probabilities of two-sided confidence intervals for log-odds ratios from a case-control study design with larger effects. The empirical coverage is calculated for increasing values of $v_2$ and for $K \in \{35, 40, 45, 50, 100, 200\}$. The curves correspond to the proposed penalized likelihood method (PLR), the DerSimonian & Laird method (DL; left column), the Zeng & Lin double resampling method (DR; left column), the Skovgaard statistic (Skov; left column), the Wald statistic (Wald; right column), the likelihood-ratio statistic (LR; right column), and the Bartlett-corrected likelihood-ratio statistic (bLR; right column). The grey horizontal line is the target 95% nominal level.



\end{document}